\documentclass[aps,prl,twocolumn,showpacs,superscriptaddress,groupedaddress,nofootinbib,reprint]{revtex4-1}
\pdfoutput=1
\usepackage[T1]{fontenc} 
\usepackage{amsmath,amssymb,graphicx,bbm,mathrsfs,nicefrac}
\usepackage{tikz}
\usepackage{amsmath}
\usepackage{amssymb}
\usepackage{graphicx,textcomp,float,gensymb,wrapfig, enumitem,comment,dsfont,framed,slashed,appendix,wrapfig,xcolor}
\usepackage{ bbold }
\usepackage{braket} 
\usepackage{ mathrsfs }
\usepackage[small]{caption}
\usepackage{subcaption}
\usepackage{etoolbox,hyperref,makecell}
\usepackage[utf8]{inputenc}
\usepackage{feynmp}
\usetikzlibrary{arrows}
\usepackage{empheq}
\DeclareMathOperator{\Tr}{Tr}

\patchcmd{\abstract}{\null\vfil}{}{}{}
\newcommand{\bea}{\begin{eqnarray}}  
\newcommand{\eea}{\end{eqnarray}}

\begin{document}

\title{Composite Higgs models after Run2}

\author{Veronica Sanz and Jack Setford} 
%\email{V.Sanz@sussex.ac.uk}
%\email{J.Setford@sussex.ac.uk}
\affiliation{Department of Physics and Astronomy, University of Sussex, 
Brighton BN1 9QH, UK}

\date{\today}% It is always \today, today,

\begin{abstract}
We assess the status of models in which the Higgs is a composite pseudo-Nambu Goldstone boson, in the light of the latest 13 TeV Run 2 Higgs data. Drawing from the extensive Composite Higgs literature, we collect together predictions for the modified couplings of the Higgs, in particular examining the different predictions for $\kappa_V$ and $\kappa_F$. Despite the variety and increasing complexity of models on the market, we point out that many independent models make identical predictions for these couplings. We then look into further corrections induced by tree-level effects such as mass-mixing and singlet VEVs. We then investigate the compatibility of different models with the data, combining the Run 1 and recent Run 2 LHC data. We obtain a robust limit on the scale $f$ of 600 GeV, with stronger limits for different choices of fermion embeddings. We also discuss how a deficit in a Higgs channel could pinpoint the type of Composite Higgs model responsible for it. 
\end{abstract}

\maketitle

\section{Introduction}

Composite Higgs models \cite{Kaplan:1983fs, Kaplan:1983sm, Agashe:2004rs} offer an elegant solution to the hierarchy problem of Higgs physics. They postulate the existence of a new strongly interacting sector which confines not far above the electroweak scale. In recent years there has been significant interest in a specific class of these models -- models in which the Higgs emerges as a pseudo-Nambu Goldstone boson of the strong sector. This sector is taken to be endowed with a global symmetry which is spontaneously broken in the confining phase, protecting the Higgs mass from corrections above the compositeness scale. Although the idea is reasonably straightforward, there are, as with most theories Beyond the Standard Model, many possibilities for its realisation.

Although this plethora of models offers a variety of unique and interesting predictions, those that are most immediately testable are the modifications of the Higgs couplings to the rest of the Standard Model fields. Of particular interest are the values of the coupling modifiers $\kappa_V$ and $\kappa_F$, as defined in \cite{Khachatryan:2016vau}.

In this paper we summarise the predictions for these couplings in Composite Higgs (CH) models. We make the case that, despite the diversity of models in the literature, these predictions have very generic structures, and we attempt to provide some intuition for this fact.

We then investigate some simple cases in which tree-level effects can modify these generic structures. These can occur, for instance, in models with extra singlets that get vacuum expectation values (VEVs), or models with an extra $SU(2)_L$ doublet that mixes with the Higgs. We point out that to leading order the modifications to $\kappa_V$ and $\kappa_F$ are precisely as one would expect in corresponding models where all the scalars are elementary, plus the usual CH corrections.

Taking the generic structures we have identified, we then perform a $\chi^2$ fit to the data, allowing for the possibility that different fermions couple in different ways. We place bounds on the compositeness scale $f$, and identify the classes of models that are most constrained. 

\section{The non-linear Composite Higgs}
\label{tnlch}

In Composite Higgs models, the Higgs is realised as a pseudo-Nambu Goldstone boson (pNGB) of a broken global symmetry. This symmetry is a symmetry of a \emph{new} strongly interacting sector, out of which the Higgs emerges as a composite.

Let the global symmetry be denoted $\mathcal G$ and the subgroup to which it spontaneously breaks be denoted $\mathcal H$. Then the Higgs and the other pNGBs (denoted collectively by $\phi^a$, one for each broken generator $X^a$), are parametrised via
\begin{equation}
U = \exp(i\phi^a X^a / f),
\end{equation}
where $f$ is an energy scale associated with the spontaneous symmetry breaking. $U$ transforms non-linearly under the global symmetry $\mathcal G$:
\begin{equation}
U \rightarrow g U h^{-1},
\end{equation}
where $g \in \mathcal G$ and $h \in \mathcal H$. By non-linear we mean that the transformation $h$ is field-dependent: $h = h(g, \phi^a)$. 

In cases where the coset $\mathcal G / \mathcal H$ is symmetric\footnote{If $T^a$ and $X^a$ are the unbroken and broken generators respectively, then the Lie algebra of a symmetric coset obeys the schematic relations$$[T,T]\sim T,\;\;[X,X]\sim T,\;\;[T,X]\sim X.$$} we are allowed to construct an object (which we will label $\Sigma$) whose transformation under $\mathcal G$ is \emph{linear}. In all the models considered here \cite{Ferretti:2013kya, Gripaios09, Ferretti14, Contino10, Barnard13, Contino:2007zz, Agashe:2004rs, Carena:2014ria, Carmona:2014iwa, Contino:2006tv, DeCurtis:2016tsm, DeCurtis:2016scv, Mrazek:2011iu, Bertuzzo:2012ya, Sanz:2015sua, Azatov:2011qy, Csaki:2017cep, Low:2015nqa, Barbieri:2015lqa}, and in the vast majority of models in the literature, $\mathcal G/\mathcal H$ will be symmetric. This reduces the task of writing down a low-energy effective theory for the pNGBs to a relatively trivial search for invariant combinations of $\Sigma$ and the other relevant fields.

We will assume that the Higgs boson is a doublet under $SU(2)_L$, which, along with $U(1)_Y$, must be embedded as an unbroken subgroup of $\mathcal G$. Although data strongly supports the doublet scenario (e.g. see LHC constraints on the ratio of couplings to $W$ and $Z$ bosons~\cite{Khachatryan:2016vau}), non-linear models have been studied in which the four scalar fields are actually a singlet and a triplet under $SU(2)_L$ \cite{Alonso:2012px, Buchalla:2013rka, Brivio:2013pma, Brivio:2016fzo}~\footnote{Note, though, that one could assume a custodially symmetric strong sector as in Ref.~\cite{Buchalla:2014eca,Krause:2016uhw}.}.

\subsection{Gauge couplings}

The couplings of the Higgs to the gauge bosons come from the kinetic term for $\Sigma$, which in the CCWZ prescription \cite{Callan69} is:
\begin{equation}
\label{kinetic}
\mathcal L_\mathit{kinetic} = \frac{f^2}{4} \Tr[D_\mu \Sigma^\dagger D^\mu \Sigma],
\end{equation}
where $D_\mu = \partial_\mu - ig A_\mu$, with $A_\mu = A_\mu^a T^a$ for each gauged generator $T^a$. We assume that the Higgs is embedded in a bidoublet $(\bf 2, 2)$ of a custodial $SO(4) \simeq SU(2)_L \times SU(2)_R \in \mathcal H$ -- this is necessary in order to protect the $\rho$ parameter from unwanted corrections \cite{Sikivie:1980hm}. Note that this imposes the non-trivial requirement that $\mathcal H$ must contain an unbroken factor of $SO(4)$.

Since we are interested in the couplings of the physical Higgs boson to SM fields, we will expand $\Sigma$ along the direction in which the Higgs will get a VEV, and set all other pNGB fields to zero. The term in \eqref{kinetic} will generically\footnote{In unusual cases the coupling may be proportional instead to $\sin^2(H/(2f))$, but all this amounts to is a redefinition of $\xi$ and an effective rescaling of $f$.} lead to a Higgs-gauge coupling of the form:
\begin{equation}
\label{gauge_coupling}
g^2 f^2 A_\mu A^\mu \sin^2(H/f),
\end{equation}
which is valid as a series expansion around $H/f$.

Expanding around the Higgs VEV $H \rightarrow \langle H \rangle + h$ (where $h$ is the physical exictation of the Higgs field) we find the gauge boson masses and couplings:
\begin{multline}
\label{expand_around}
\mathcal L_\mathit{gauge} \supset \frac{1}{8} g^2 f^2 \sin^2\left(\frac{\langle H \rangle}{f}\right) W_\mu^a W^{a\mu} \\
+ \frac{1}{8} g^2 f \sin\left(\frac{2\langle H\rangle}{f}\right) W_\mu^a W^{a\mu} h \\ 
+ \frac{1}{8} g^2 \cos\left(\frac{2\langle H\rangle}{f}\right) W_\mu^a W^{a\mu} h^2.
\end{multline}
Identifying\footnote{Here $v$ is defined as $4M_W^2/g^2$, as in the Standard Model} $v = f \sin(\langle H \rangle / f)$ and defining $\xi = v^2/f^2$, we find
\begin{multline}
\mathcal L_\mathit{gauge} \supset \frac{1}{8}g^2 v^2 W_\mu^a W^{a\mu} \\
+ \frac{1}{4} g^2 v \sqrt{1 - \xi} W_\mu^a W^{a\mu} h + \frac{1}{8} g^2 (1 - 2\xi) W_\mu^a W^{a\mu} h^2.
\end{multline}
Thus
\begin{align}
\begin{split}
g_{WWh} &= \sqrt{1-\xi} g_{WWh}^{SM} \\
g_{WWhh} &= (1-2\xi) g_{WWhh}^{SM}.
\end{split}
\end{align}
Since $\kappa_V$ is defined as $g_{WWh}/g_{WWh}^{SM}$, we find 
\begin{equation}
\kappa_V = \sqrt{1-\xi} \approx 1 - \frac{1}{2}\xi
\end{equation}

Since the structure of \eqref{kinetic} is generic, so too is this result, at leading order, across all Composite Higgs models.

\subsection{Fermion couplings}

In Composite Higgs models the SM fermions usually couple to the strong sector via the \emph{partial compositeness} mechanism \cite{Kaplan91, Contino06, Contino10}. As far as this mechanism pertains to the construction of the low energy effective theory, it involves embedding the SM fermions in representations of the global symmetry $\mathcal G$, and then constructing $\mathcal G$ invariant operators out of these multiplets and $\Sigma$. Such an embedding is sometimes called a spurion -- the term spurion refers to the `missing' elements of the multiplet, since after all, the SM particles do not come in full multiplets of the new symmetry $\mathcal G$. The incompleteness of these spurious multiplets contributes to the explicit breaking of $\mathcal G$ and allows the Higgs to acquire a potential via loops of SM fermions.

The appropriate representation in which to embed the SM particles would, in principle, depend on the UV completion of the model. Some attempts towards UV completions of Composite Higgs models have been made (see, for example \cite{Ferretti:2013kya, Ferretti14, Barnard13}), however for the purposes of most model building the choice of representation is a `free parameter' of the model. There is, however, good cause to restrict the choice of representation into which the $SU(2)_L$ quark doublet is embedded. As shown in \cite{Agashe06}, embedding $q_L$ into a bidoublet $(\bf 2, 2)$ of the custodial $SO(4) \simeq SU(2)_L \times SU(2)_R$ can prevent anomalous contributions to the $Z\rightarrow b\overline b$ coupling. This restriction forces one to choose representations that contain a bidoublet in their decomposition under the custodial $SO(4)$ subgroup of $\mathcal G$.

To treat the EFT in full generality, one should embed $q_L$, $t_R$ and $b_R$ into different multiplets $\Psi_q$, $\Psi_t$ and $\Psi_b$. The kind of representation that the three quarks are embedded into need not be the same. Thus, even for each coset $\mathcal G/\mathcal H$, there are a bewildering number of possibilities. However, for the vast majority of models the form of $\kappa_F$ is actually quite restricted. We tabulate a few examples in Table~\ref{kappa_F}.

\begin{table}[t]
\begin{tabular}{| c | c |}
\hline
$\kappa_F$ & Models \\
\hline
$\kappa_F^A = \sqrt{1-\xi}$ & $SO(5)/SO(4)$ -- \cite{Agashe:2004rs, Carena:2014ria} \\
& $SO(6)/SO(4)\times SO(2)$ -- \cite{Mrazek:2011iu, DeCurtis:2016tsm, DeCurtis:2016scv} \\
& $SU(5)/SU(4)$ -- \cite{Bertuzzo:2012ya} \\
& $SO(8)/SO(7)$ -- \cite{Low:2015nqa, Barbieri:2015lqa} \\
\hline
$\kappa_F^B = \frac{1-2\xi}{\sqrt{1-\xi}}$ & $SO(5)/SO(4)$ -- \cite{Contino:2006tv, Carena:2014ria, Carmona:2014iwa, Csaki:2017cep} \\
 & $SU(4)/Sp(4)$ -- \cite{Gripaios09} \\
 & $SU(5)/SO(5)$ -- \cite{Ferretti14} \\
 & $SO(6)/SO(4) \times SO(2)$ -- \cite{Mrazek:2011iu, DeCurtis:2016tsm, DeCurtis:2016scv} \\
 \hline
\end{tabular}
\caption{$\kappa_F$ in different models.}
\label{kappa_F}
\end{table}

It might seem strange that so many distinct models lead to so few possibilities for $\kappa_F$. In fact, when one examines the structure of the allowed terms in the effective Lagrangian, a general pattern emerges: the lowest order coupling of the Higgs to fermions will generally contain either one or two factors of $\Sigma$. For example, in the Minimal Composite Higgs Model (MCHM), the coset group is $SO(5)/SO(4)$, and one can define a linearly transforming $\Sigma$ in the $\bf 5$ of $SO(5)$, which, expanded along the $H$ direction can be expressed as
\begin{equation}
\Sigma(h) = (0,0,0,\sin(H/f),\cos(H/f)).
\end{equation}
With $q_L$ and $t_L$ embedded in the $\bf 5$, Yukawa couplings come from the $SO(5)$ invariant effective operator
\begin{equation}
(\overline\Psi_q^{\bf 5} \cdot \Sigma) (\Sigma \cdot \Psi_t^{\bf 5}),
\end{equation}
leading to a term proportional to $\sin(H/f)\cos(H/f)$. Alternatively one could embed $q_L$ into a $\bf 10$, the $t_R$ into a $\bf 5$ -- in this case the Yukawa term originates from an operator like
\begin{equation}
\Sigma^T \overline\Psi_q^{\bf 10} \Psi_t^{\bf 5},
\end{equation}
and the interaction is proportional to $\sin(H/f)$~\footnote{Note that this structure of couplings also depends on the assumption that the Higgs forms part of a doublet, whereas other forms of the effective coupling could be possible in a singlet case, see e.g. the generic forms of the potential in Ref.~\cite{Croon:2015fza}.}.

In general the structure must be such that the leading term in the trigonometric expansion is $H/f$. In almost all cases the relevant term will be proportional to either $\sin(H/f)$ or $\sin(H/f)\cos(H/f)$. This argument is certainly not intended to be rigorous -- we merely hope to provide some intuition for the fact the non-linear nature of a pNGB Higgs boson leads to repeated structures even across different models and choices of representations\footnote{See also \cite{Bellazzini:2014yua} for a comprehensive review of different Composite Higgs models, and an especially detailed look at the constraints on the $SO(5)/SO(4)$ coset with Run 1 data.}.

Following the same procedure as in equation \eqref{expand_around}, we can expand around the Higgs VEV to find the expression for $\kappa_F$, defined by $y v / m_F$. A coupling of the form $\overline \psi \psi \sin(H/f)$ leads to
\begin{equation}
\kappa_F = \sqrt{1 - \xi} \approx 1-\frac{1}{2}\xi,
\end{equation}
while a coupling of the form $\overline \psi \psi \sin(H/f) \cos(H/f)$ leads to
\begin{equation}
\kappa_F = \frac{1-2\xi}{\sqrt{1-\xi}} \approx 1-\frac{3}{2}\xi.
\end{equation}

As we stated above, the representation into which we embed $t_R$ and $b_R$ might not be the same -- in this case it is quite possible (depending on the details of the model) that the top and bottom couplings to the Higgs have different structures. For instance, in the second example above, although the $t_R$ is embedded into a $\bf 5$, the $b_R$ might be embedded into a $\bf 10$. As a result the top coupling would scale with $1-\frac{1}{2}\xi$ while the bottom coupling would scale with $1-\frac{3}{2}\xi$.

There are (as always) some interesting exceptions. For example, in \cite{Azatov:2011qy}, with $q_L$ in a $\bf 5$ and $t_R$ in a $\bf 14$, one can derive $\kappa_F \approx 1 - 3\xi$, see also Ref.~\cite{Montull:2013mla}. In some models (for some examples, see \cite{Carena:2014ria, Azatov:2011qy}) more than one operator can be constructed which contributes to the same Yukawa coupling. The degree to which each operator contributes will, in such cases, be a free parameter and will lead to more complex expressions for $\kappa_F$. Such models are interesting insofar as they are exceptions -- however more minimal scenarios will follow the structure we have outlined above.

No mention has been made so far of the leptonic sector. In theory the lepton Yukawas can also be generated via the partial compositeness mechanism (see for instance \cite{Carmona:2014iwa}). This means that $\kappa_\tau$ (for instance) would also receive corrections, and in minimal scenarios would depend on $\xi$ like $\kappa_F^A$ or $\kappa_F^B$, as defined in Table~\ref{kappa_F}.

\section{Tree-level effects}
\label{tle}

In this section we will briefly look at two interesting scenarios that can lead to tree-level corrections to $\kappa_V$ and $\kappa_F$ from the integrating-out of heavier states. We will describe these corrections as leading to a new effective $\xi_\mathit{eff}$ to be compared with the vanilla prediction for $\xi$.

The first possibility is that in models with an extra singlet pNGB (such as the $SU(4)/Sp(4)$ and $SU(5)/SO(5)$ cosets), the pNGB potential could induce a VEV for the singlet. This can modify $\kappa_F$ and $\kappa_V$ in two ways -- firstly a VEV for the singlet $\eta$ will induce singlet-doublet mixing between $\eta$ and $H$. Singlet-doublet mixing (in the elementary case) and its effect on Higgs couplings was studied in detail in \cite{Gorbahn:2015gxa}. The fact that the $H$ mixes with another scalar means that the couplings will be modified by a factor of $\cos\theta$, where $\theta$ is the mixing angle between $H$ and $\eta$. For small mixing angles:
\begin{equation}
\kappa_V \approx 1 - \frac{1}{2}\theta^2.
\end{equation}

In this and in the following we are assuming that the singlet is heavier than the Higgs and that it makes sense to integrate it out. Generally, in the absense of further tuning, one expects the extra pNGBs to be heavier than the Higgs by a factor of $\xi = v^2/f^2$, since this is the amount by which the mass of the Higgs has to be tuned to satisfy electroweak precision test \cite{Franceschini:2015kwy}. Thus, in models with around 10\% tuning, values for the extra pNGB masses of around $300-500$ GeV are not unreasonable.

There could also be effects similar to those studied above, arising from higher-dimensional terms in the non-linear effective theory. As an example we will look at the $SU(4)/Sp(4)$ model. The gauge boson coupling to the Higgs and $\eta$ (the equivalent of equation \eqref{gauge_coupling}) will be (neglecting hypercharge)
\begin{equation}
\frac{H^2}{H^2 + \eta^2} \sin^2\left(\frac{\sqrt{H^2 + \eta^2}}{f}\right) W^a_\mu W^{a\mu}.
\end{equation}
As expected, there is no dimension-4 coupling of $\eta$ to the $SU(2)_L$ gauge bosons, but there are higher order terms involving $\eta$ which could modify the $hWW$ coupling if $\eta$ gets a VEV. However one should also note that the kinetic term in \eqref{kinetic} corrects the Higgs kinetic term:
\begin{equation}
\mathcal L_\mathit{kinetic} = \frac{\sin^2(v_\eta/f)}{v_\eta^2/f^2} (\partial_\mu H)^2 \approx (1 - \frac{1}{3}\xi_\eta) (\partial_\mu H)^2.
\end{equation}
After canonically normalising the Higgs field and expanding around small values of $\xi_\eta = v_\eta^2/f^2$ we find that the $\mathcal O(\xi_\eta)$ correction to $\kappa_V$ actually cancels. To leading order in $\xi, \xi_\eta$ and $\theta$ we have:
\begin{equation}
\kappa_V \approx 1 - \frac{1}{2}\xi - \frac{1}{2}\theta^2.
\end{equation}
The correction due to the singlet VEV thus neatly ``factorises'' into the mass-mixing correction $\mathcal O(\theta^2)$ plus the usual compositeness correction $\mathcal O(\xi)$. We can thus define a $\xi_\mathit{eff} = \xi + \theta^2$, such that $\kappa_V \approx 1-\xi_\mathit{eff}/2$.

One finds a similar result for $\kappa_F$. The singlet VEV modifies $\kappa_F$ from $\approx 1-\frac{3}{2}\xi$ to
\begin{equation}
\kappa_F \approx 1 - \frac{3}{2}\xi - \frac{1}{2}\theta^2,
\end{equation}
and in this case our effective $\xi_\mathit{eff} = \xi + \frac{1}{3}\theta^2$.

In the regime where $m_\eta$ and $v$ are both $\gg v$, the mixing will be small and will scale approximately as
\begin{equation}
\theta^2 \sim \frac{v^2 v_\eta^2}{m_\eta^4} = \frac{1}{g_\eta^4} \xi \xi_\eta,
\end{equation}
where we have related $m_\eta$ to $f$ via some coupling: $m_\eta = g_\eta f$.

The amount of tuning present in such a model was analysed in \cite{Banerjee:2017qod}. This coset was also investigated in a cosmological setting in \cite{Croon:2015fza,Croon:2015naa}, where the singlet $\eta$ plays the role of the inflaton. In such a scenario the size of the singlet VEV has important implications for the scale of inflation, and the mass-mixing of the inflaton would be important also for the process of reheating. Moreover, the singlet $\eta$ and a non-zero value of $\xi_\eta$ could be a key component of a solution to the matter-antimatter asymmetry in the Universe~\cite{Espinosa:2011eu}.

If the value of $\xi_\mathit{eff}$ were the same for all couplings (i.e. the modifications to $\kappa_V$ and $\kappa_{F_i}$ were the same), then the theory would resemble a CH model without any mixing, only with an apparent rescaling of $f$. However it is interesting to note that in the above case the inferred values of $\xi_\mathit{eff}$ from the measurements of $\kappa_V$ and $\kappa_F$ are different, which would in principle allow us to experimentally distinguish between these two scenarios.

Another possibility is that the spontaneous breaking leads to another pNGB doublet of $SU(2)_L$ (a composite two Higgs doublet model). In principle, explicit breaking effects could lead to a mixing between the two doublets. This possibility is discussed in \cite{DeCurtis:2016tsm, DeCurtis:2016scv}, and in a different context in \cite{Sanz:2015sua}, in which the two doublets appear from two different spontaneous breakings at different scales.

In this case we will obtain similar results to our expressions above for $\xi_\mathit{eff}$, with a correction from the mass-mixing at $\mathcal O(\theta^2)$ that will be present in the elementary case, and the usual correction at $\mathcal O(\xi)$ coming from higher dimensional operators (see \cite{Branco:2011iw} for a review of the elementary two Higgs doublet model, and \cite{Gorbahn:2015gxa} for an analysis of the Higgs EFT in such a scenario).

Since we have looked at tree-level corrections to $\kappa_V$ and $\kappa_F$ coming from new states in the composite sector, one should also talk about loop level modifications. In principle loops of scalar, fermionic and vector resonances of the strong sector can modify the Higgs couplings. These will arise from higher dimensional ($d \ge 6$) operators in the effective theory, suppressed by factors of $f^{4-d}$.

\section{Status after Run 2}
\label{sar2}
 
\begin{figure*}[t!]%
    \begin{center}
  \includegraphics[scale = 0.65]{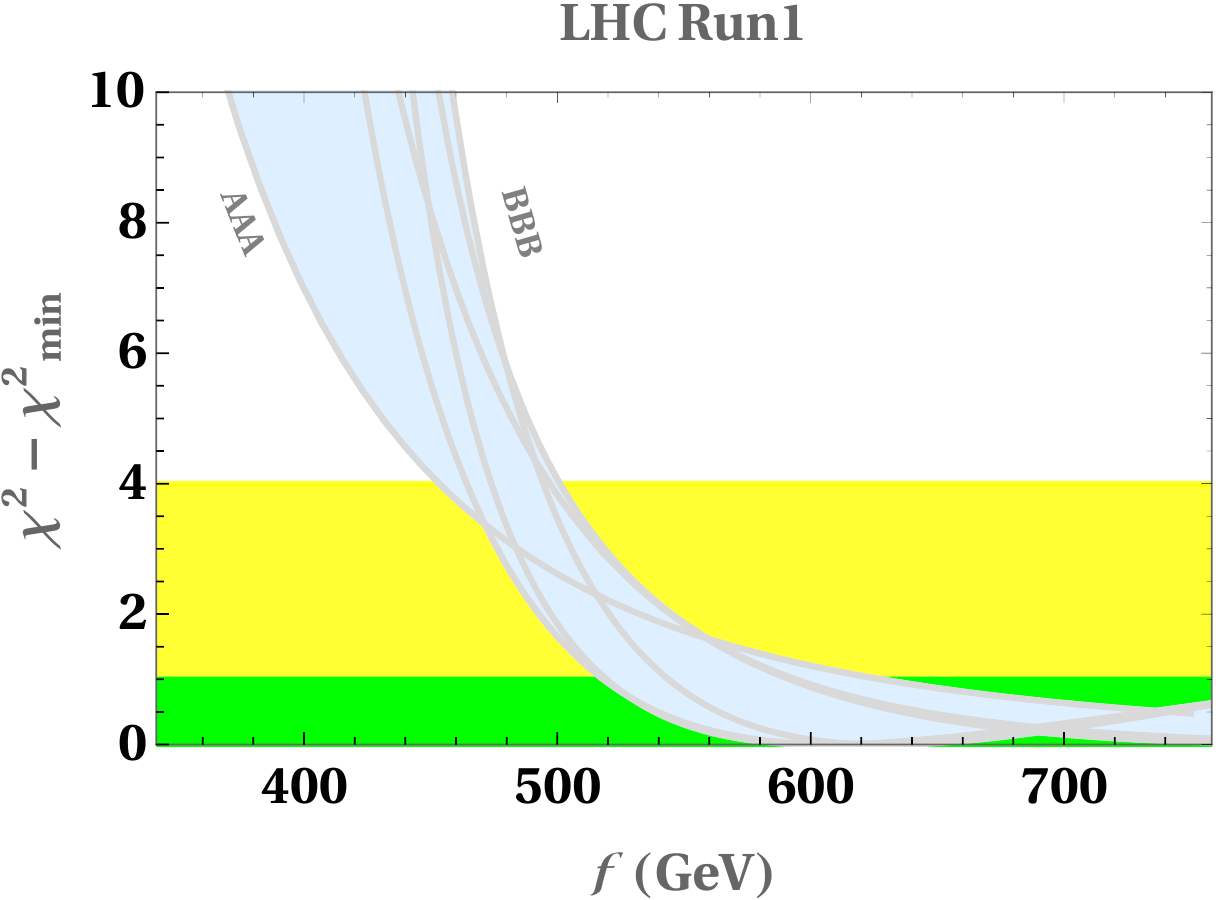}
  \hspace{7mm}
  \includegraphics[scale = 0.65]{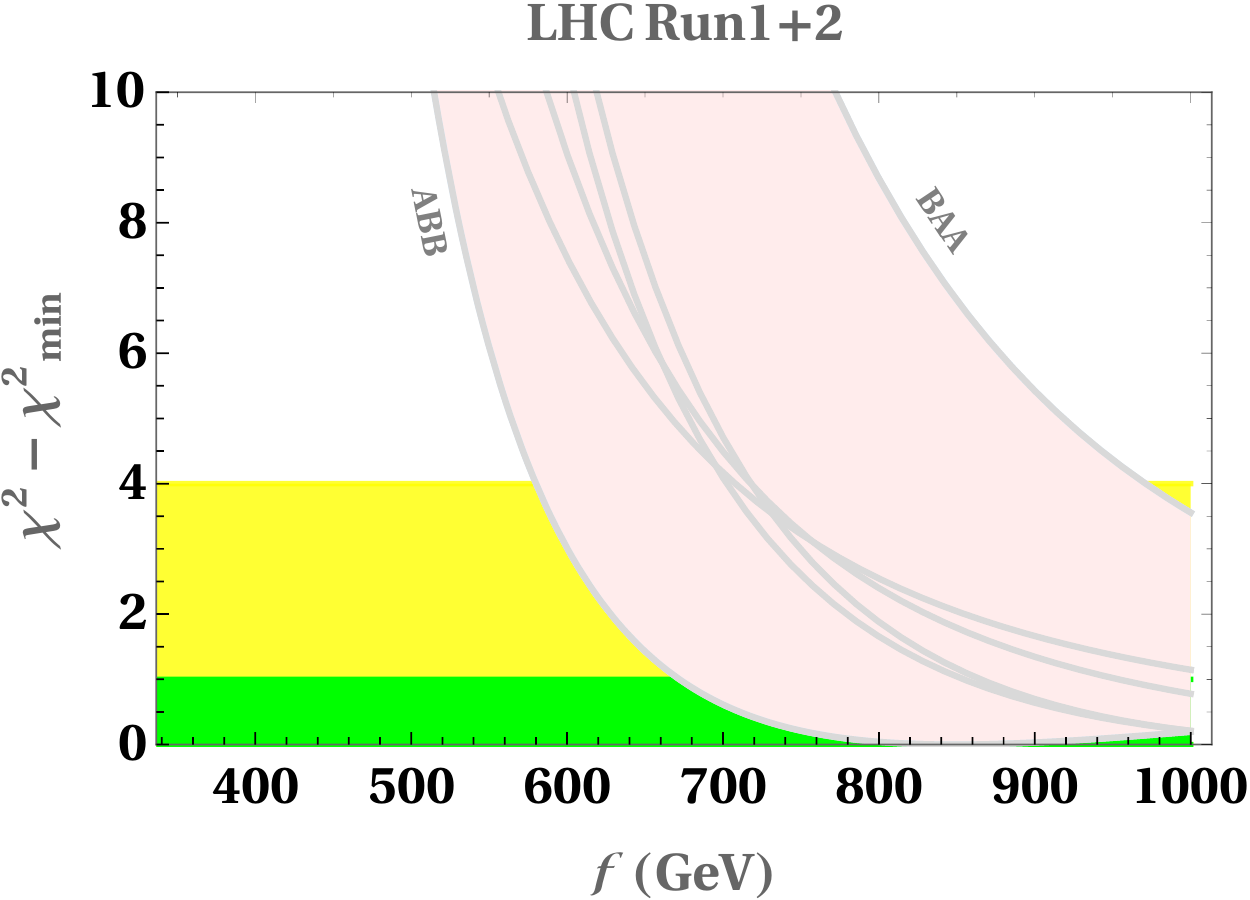}
    \end{center}
    \caption{$\chi(f)^2-\chi^2_{min}$ for the Run 1 (left) and combination of Run 1 and 2 (right) datasets. The lines correspond to different choices of fermion couplings $\kappa_F^{A,B}$ for $(\kappa_t,\kappa_b,\kappa_\tau)$. For example, AAA indicates $\kappa_t=\kappa_b=\kappa_\tau=\kappa_F^A$.  }\label{fig:results}
\end{figure*} 

In this section we study the impact of Run 1 LHC data on Composite Higgs models, as well as the improvement which results when adding the 13 TeV results recently released by the collaborations. In Table~\ref{table} we summarize the channels considered in the combination of Run 1 and 2 data from ATLAS and CMS, as well as indicate the coupling modifiers that one would obtain in Composite Higgs models, as discussed previously. 

\begin{table}[h!]
\centering
\begin{tabular}{|c|c|c|}
\hline
 & & \\
Channel & Refs. & $\kappa$-factors\\ & & \\
\hline
$ttH$ ($H \to \gamma\gamma$) & \cite{Aaboud:2017jvq, ATLAS:2017myr, CMS:2017rli} & $\frac{\kappa_t^2 \kappa_\gamma^2}{\kappa_H^2}$ \\
$ttH$ ($H \to b \bar b$) & \cite{Aaboud:2017jvq} & $\frac{\kappa_t^2 \kappa_b^2}{\kappa_H^2}$ \\
$ttH$ ($H \to \tau^+ \tau^-$) & \cite{Aaboud:2017jvq} & $\frac{\kappa_t^2 \kappa_\tau^2}{\kappa_H^2}$ \\
$ttH$ ($H \to W W^*$, $H \to Z Z^*$) & \cite{Aaboud:2017jvq} & $\frac{\kappa_t^2 \kappa_V^2}{\kappa_H^2}$ \\
$ggF$ ($H \to \gamma\gamma$) & \cite{ATLAS:2017myr, CMS:2017rli} & $\frac{\kappa_g^2 \kappa_\gamma^2}{\kappa_H}$ \\
$ggF$ ($H \to \tau^+ \tau^-$) & \cite{CMS:2017wyg} & $\frac{\kappa_g^2 \kappa_\tau^2}{\kappa_H^2}$ \\
$ggF$ ($H \to W W^*$, $H \to Z Z^*$) & \cite{CMS:2017pzi, Aaboud:2017vzb, CMS:2017jkd} & $\frac{\kappa_g^2 \kappa_Z^2}{\kappa_H^2}$ \\
$HV$ ($H \to b \bar b$) & \cite{Aaboud:2017xsd, Sirunyan:2017elk} & $\frac{\kappa_V^2 \kappa_b^2}{\kappa_H^2}$ \\
$VBF$, $HV$ ($H \to \gamma\gamma$) & \cite{ATLAS:2017myr, CMS:2017rli} & $\frac{\kappa_V^2 \kappa_\gamma^2}{\kappa_H}$ \\
$VBF$, $HV$ ($H \to W W^*$, $H \to Z Z^*$) & \cite{ATLAS:2016gld, CMS:2017pzi, CMS:2017jkd} & $\frac{\kappa_V^4}{\kappa_H^2}$ \\

\hline
\end{tabular}
\caption{\label{tab:pheno} List of 13 TeV channels considered in the fit, with the corresponding $\kappa$ modifyiers. Note that the 7+8 TeV Run 1 data was included using the results of the combination of ATLAS and CMS data in Ref.~\cite{Khachatryan:2016vau}.} \label{table}
\end{table}

The couplings of the Composite Higgs to gluons and photons,  $\kappa_g$ and $\kappa_\gamma$, are functions of the modifications of the couplings to fermions and gauge bosons, which appear at one-loop order, i.e. $\kappa_g^2=1.06 \kappa_t^2+0.01 \kappa_b^2-0.07 \kappa_b \kappa_t$ and  $\kappa_\gamma^2=1.59 \kappa_V^2 +0.07 \kappa_t^2-0.66 \kappa_V \kappa_t$~\cite{Khachatryan:2016vau,Gillioz:2012se}. The modification of the Higgs width, $\kappa_H$ is also a function of the coupling modifiers, $\kappa_H^2 \approx 0.57 \kappa_b^2 + 0.25 \kappa_V^2 + 0.09 \kappa_g^2$, see e.g. Ref.~\cite{Khachatryan:2016vau}. 

We then perform a $\chi^2$ fit to the ATLAS and CMS data\footnote{When two measurements of the same channel were available, we discarded the worse measurement, or kept both if they were of similar significance. Results from \cite{ATLAS:2016lgh, CMS:2017vru} were considered but not included in the fit.}, with the restriction $\xi >0$.

The dependence of the $\chi^2$ function with the scale of new physics $f$ is shown in Fig.~\ref{fig:results}. The green and yellow bands correspond to the one- and two-sigma regions of the fit, and the left and right panels correspond to Run 1 and the combination of Run 1 and Run 2, respectively. Different choices of fermion representations $\kappa_F^{A,B}$ (as shown in Table~\ref{kappa_F}) lead to different $\chi^2$ dependences.

\begin{figure}%
    \begin{center}
  \includegraphics[scale = 0.65]{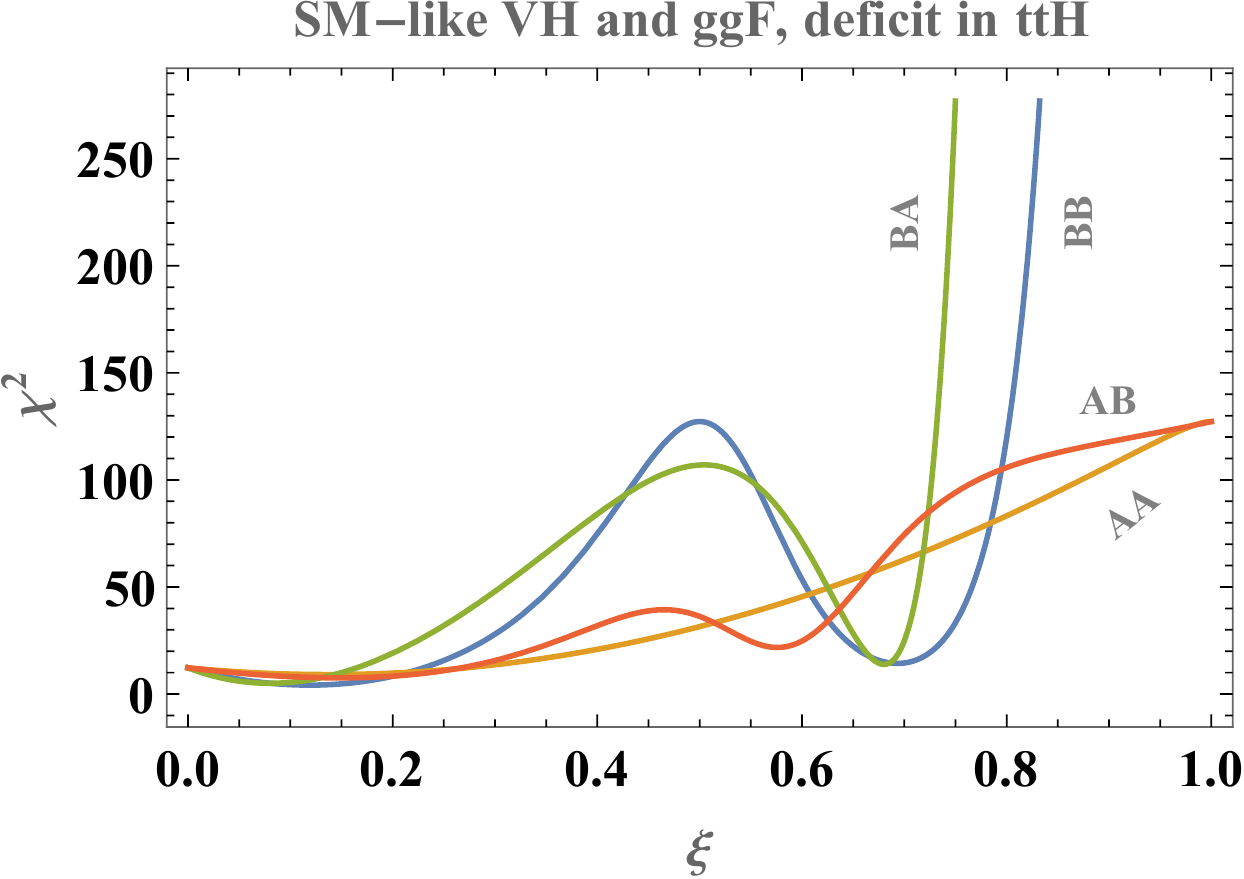}
     \end{center}
    \caption{$\chi(\xi)^2$ assuming a scenario where a deficit is found in $ttH$ production channels, while other channels remain consistent with the SM. The labels correspond to different hypothesis of $\kappa_F^{A,B}$ for ($\kappa_t$, $\kappa_b$). In this case, the choice $\kappa_t=\kappa_b=\kappa_F^B$ would be preferred by data. We assume a 20\% uncertainty in these channels, except in $gg \to H \to \gamma\gamma$ where  a 10\% accuracy is assumed. }\label{fig:future}
\end{figure}

The  model-independent limit on $f$ improves from 450 GeV (Run 1) to 600 GeV (Run 1+2) at 95\% CL, and we see that the most constrained screnario is $\kappa_t = \kappa_F^A$, $\kappa_b = \kappa_\tau = \kappa_F^B$. Moreover, one can see that the spread of limits on the scale $f$ due to these fermion choices increases with the addition of more data. This is a signal that the data is increasingly sensitive to these choices, due to better determination of the Higgs couplings to the heavy fermions. To illustrate this point, assume that at some point in the future a deficit in one channel is observed, whereas other channels remain consistent with the SM. For example, assume that the signal strength of the $ttH$ processes was found to be a third of the SM rate, whereas other processes involving the coupling of the Higgs to vector bosons remained consistent with the SM. In this case, certain representations for fermion embeddings of the top and bottom quarks would be preferred by data, see Fig.~\ref{fig:future}.

These limits on $f$ should be compared with the limits of direct searches for new resonances. One would typically expect a set of new resonances, e.g. new massive $W'$ and $Z'$, to appear at some scale related to $f$, $m_{W'} = g_\rho f$, with $g_\rho \lesssim {\cal O}(4\pi)$. The value of $g_\rho$ is an input to the effective theory, but can be obtained by performing a lattice simulation of the theory and investigating the spectrum of resonances.  Its value depends on the specific pattern of breaking as well as the possible electroweak effects. As an indicator of the value of $g_\rho$ in these kind of models, we draw attention to the work done in the coset $SO(6)/SO(5)$~\cite{Lewis:2011zb}, and in others scenarios \cite{Appelquist:2016viq}, where $g_\rho$ was found to be ${\cal O}(10)$. In this case, a limit on $f\sim$ 600 GeV, would correspond to a $Z'$ and $W'$ in the multi-TeV scale, certainly competitive with direct searches for these resonances.   

Besides vector resonances, one would expect a tower of fermion resonances, or techni-baryons. Typically, these techni-baryons are heavier than the vector bound states by a factor of $N_{TC}$, with $N_{TC}$ the number of colours in the new strongly coupled sector~\cite{tHooft:1973alw,Witten:1979kh}. Hence, naively one would expect fermion resonances again in the multi-TeV scale. Yet, in most Composite Higgs models the mechanism of electroweak symmetry breaking depends on the existence of light techni-baryons ({\emph{top partners}) with masses of the order of $f$, contrary to the large-$N$ expectation. This mechanism is being tested by direct searches of heavy partners of the top, with recent Run 2 results already sensitive to the 1.2 TeV region~\cite{ATLAS-CONF-2017-015}, clearly more competitive than the indirect searches in Higgs data if one believed this is the correct mechanism in place. Note, though, that the mass of the top partner is also linked to the amount of fine-tuning in these models. From this point of view the strong limits in top-partners may lead one to consider alternative constructions, such as Composite Twin Higgs models \cite{Batra:2008jy, Barbieri:2015lqa, Low:2015nqa, Csaki:2015gfd}, or models involving the see-saw mechanism devloped in \cite{Sanz:2015sua}. In such models the top partners can be significantly heavier without introducing more fine-tuning.

\section{Conclusions}

In this paper we have summarised the structure of the Higgs couplings (parameterised by $\kappa_V$ and $\kappa_F$) in Composite Higgs models. Although different CH models have very different predictions for the UV theory and the spectrum of higher mass resonances, we have identified generic forms for $\kappa_V$ and $\kappa_F$ which hold for many different choices of the coset group and fermion representations.

We have also looked into tree level effects on these couplings coming from extra states. In particular we studied the interesting possibility that an extra singlet pNGB may acquire a VEV. The modifications to $\kappa_V$ and $\kappa_F$ are to leading order just a sum of the corrections in elementary singlet + doublet models, and the usual correction expected in composite models. The same can be said for the case in which the Higgs mixes with an extra doublet.

We combined the Run 1 and recent Run 2 LHC data to set limits on CH models, finding that different choices for fermion representations lead to a spread of limits but a lower bound on the scale $f$ can be set to 600 GeV. We also discussed how an observed deficit in a Higgs channel such as $t \bar t H$ could pinpoint the type of CH model responsible for it.

\section{Acknowledgements}

This work is supported by the Science Technology and Facilities Council (STFC) under grant number ST/J000477/1.

\bibliographystyle{unsrt}
\bibliography{References}

\end{document}